 \documentclass[sigconf]{acmart}

\AtBeginDocument{%
  \providecommand\BibTeX{{%
    \normalfont B\kern-0.5em{\scshape i\kern-0.25em b}\kern-0.8em\TeX}}}

\setcopyright{none}
\acmDOI{}
\acmISBN{}
\acmConference[Proceedings of the ImpactRS Workshop at ACM RecSys '20]{}{September 25, 2020}{Virtual Event, Brazil.}
\copyrightyear{Copyright (c) 2020 for this paper by its authors. Use permitted under Creative Commons License Attribution 4.0 International (CC BY 4.0)}

\usepackage{makecell}

\begin{document}

\title{Exploring Artist Gender Bias in Music Recommendation}


\author{Dougal Shakespeare$^1$, Lorenzo Porcaro$^1$, Emilia G\'{o}mez$^{1,2}$, Carlos Castillo$^3$}
\affiliation{%
    \institution{$^1$Music Technology Group, Universitat Pompeu Fabra, Barcelona, Spain}
}
\affiliation{%
    \institution{$^2$Joint Research Centre, European Commission, Seville, Spain}
}
\affiliation{%
    \institution{$^3$Web Science and Social Computing Group, Universitat Pompeu Fabra, Barcelona, Spain}
}
\email{dougalian.shakespeare01@estudiant.upf.edu}
\email{{lorenzo.porcaro, emilia.gomez, carlos.castillo}@upf.edu}

\renewcommand{\shortauthors}{Shakespeare, et al.}

\begin{abstract}
Music Recommender Systems (mRS) are designed to give personalised and meaningful recommendations of items (i.e. songs, playlists or artists) to a user base, thereby reflecting and further complementing individual users' specific music preferences.
Whilst accuracy metrics have been widely applied to evaluate recommendations in mRS literature, evaluating a user's item utility from other impact-oriented perspectives, including their potential for discrimination, is still a novel evaluation practice in the music domain.
In this work, we center our attention on a specific phenomenon for which we want to estimate if mRS may exacerbate its impact: \textit{gender bias}.
Our work presents an exploratory study, analyzing the extent to which commonly deployed state of the art Collaborative Filtering (CF) algorithms may act to further increase or decrease artist gender bias. 
To assess group biases introduced by CF, we deploy a recently proposed metric of bias disparity on two listening event datasets: the LFM-1b dataset, and the earlier constructed Celma's dataset.
Our work traces the causes of disparity to variations in input gender distributions and user-item preferences, highlighting the effect such configurations can have on user's gender bias after recommendation generation.
\end{abstract}



\maketitle
\pagestyle{empty}

\section{Introduction}\label{sec:introduction}
Impact-oriented Recommender System (RS) research is gaining attention as a novel paradigm for understanding not only how users interact with recommendations, but also for shedding light on how these interactions can influence users' behaviours in the short- and the long-term \cite{Jannach2019}.
An outstanding issue when studying the possible impact of RS is the heterogeneity of evaluation procedures described in the literature.
Evaluating recommender systems is a non-trivial task because of the multiple facets that a \textit{good} recommendation can have, and the multiple players influencing these aspects \cite{Gunawardana2015}.
Even if the need for going beyond the evaluation in terms of accuracy metrics has been well-recognized by the RS community \cite{McNee06}, shared practices for evaluating the impact of recommendations still are missing.

Notwithstanding, recent years have seen a rise in awareness in the scientific community about the implications of socio-technical systems' design and implementation responsible of reinforcing bias and discrimination \cite{West2019, BaezaYates2018}. Music Information Retrieval (MIR) research is still in its early-stage with regards to the analysis of the ethical dimensions and impact of music technology \cite{Serra2013, Holzapfel2018, Salamon2019, gomez2019}, and several challenges still need to be tackled when approaching MIR research from a socio-technical perspective.
A common issue is the availability of data,  often limited in terms of size, user information or musical information, and as in many other fields, a chronic shortage of gender-disaggregated data~\cite{perez2019invisible}.
The difficulties in our research to retrieve the artists' gender are just one example of this limitation, as presented in Section 3 and 4.

We center our attention on a specific phenomenon that recommender systems may exacerbate: \textit{gender bias}.
In its broader sense, gender discrimination is a disadvantage for a group of people based on their gender.
Far from being an emerging problem, gender discrimination has its roots in cultural practices historically related with socio-political power differentials \cite{Beauvoir2015}.
Nonetheless, the modern day prevalence of gender discrimination is not to be understated: recent reports find the disproportionate treatment of female artists to be prevalent in the Western music industry to this day \footnote{\url{http://assets.uscannenberg.org/docs/aii-inclusion-recording-studio-2019.pdf}}. Whilst the cause of such treatment is multifaceted, our work traces the influence of one factor evidenced to be present in the works of Millar \cite{Millar2008} that is, the pre-existing gender bias of a music listener.

In this exploratory study, we assess the extent to which Collaborative Filtering (CF) algorithms commonly deployed in mRS may exacerbate pre-existing users' gender biases thereby affecting an artist gender's exposure and proportional representation. We focus on the measurement of bias disparity in recommender systems, defined as "\textit{[...] the case where the recommender system introduces bias in the data, by amplifying existing biases and reinforcing stereotypes.}" \cite{Tsintzou2018}.
Building on existing literature \cite{Tsintzou2018, Zhao2017, Lin2019, Mansoury2019}, we first reproduce the study presented by Lin et al. \cite{Lin2019}, in which preference bias amplification in collaborative recommendation is analyzed using the MovieLens dataset\cite{Harper2015}, a dataset of user activity with a movie recommendation system.
In our work, we focus on the music domain making use of two Last.fm\footnote{\url{https://www.last.fm
}} listening event datasets publicly available: 1) Celma's LFM-360k dataset \cite{Celma2010}; 2) Schedl's LFM-1b dataset \cite{Schedl2016}.
Our goal is twofold: on one hand, reproducing and verifying whether previous results \cite{Lin2019} hold across different datasets.
On the other hand, we aim at highlighting which aspects specific to the music domain can be extracted by this analysis, connecting with existing literature on gender bias in music preferences \cite{Millar2008, AngladaTort2019}.

The paper is structured as follows. Section \ref{sec:relatedwork} provides an overview of previous works related to bias in Information Technology, focusing on gender bias, but also how this bias has been approached in music-related fields.
We then introduce the considered datasets, LFM-1b and LFM-360K respectively in Section \ref{subsec:lfm1bdataset} and \ref{subsec:lfm360kdataset}. In Section \ref{sec:methodology}, the recommendation models used and the experimental settings are presented, followed by Section \ref{sec:results} which details the results obtained. Lastly, in section \ref{sec:conclusion} conclusions and future work are discussed.

\section{Related Work}
\label{sec:relatedwork}

The notion of bias has been extensively explored in the Information Retrieval domain \cite{BaezaYates2018, Barocas2014, Bozdag2013, Jannach2015, Cramer2018}. Typically, metrics aim to capture \textit{relative bias} (i.e. bias pre-existing in data, for example in user listening histories in LFM-1b), and \textit{algorithmic bias} (i.e. how filtering algorithms can result in unfair item and user treatment) to measure disproportionate unfair treatment of a protected group.

One of the most well-studied biases in RS literature is popularity bias, with the music domain being no exception to this phenomenon \cite{Celma2010, Bauer2019, Kowald2020}.
This describes the scenario in which a few popular items are recommended frequently, while the majority of items in the long-tail do not get proportional attention.
Highlighted in literature as a prominent issue for CF algorithms \cite{Celma2010, Park2008, Abdollahpouri2019}, Kowald et al. in \cite{Kowald2020} find that from a user's perspective the groups who do not favor popular items may receive worsened recommendations in terms of accuracy and calibration.
Moreover, Ferraro et al. in \cite{Ferraro2019} study the effect of musical styles with respect to popularity bias, showing that CF approaches increase users' exposure to popular musical styles.

\textit{Bias Disparity} is a metric deployed to assess bias propagation across user's and item's group, measuring the deviation of the recommender output from the input preference, as detailed in Section \ref{subsec:bias}.
A first application to the RS domain was described by Tsintzou et al.~\cite{Tsintzou2018}, but the metric has recently gained more traction in its application to different domains.
In Lin et al.~\cite{Lin2019}, bias disparity is applied to measure the extent to which state of the art CF algorithms can exacerbate pre-existing biases in the MovieLens dataset.
Their findings show significant differences in bias propagation across memory- and model-based CF algorithms.

Gender treatment and issues of proportional treatment in RS have been considered in a range of literature, for which we highlight some examples.
Ekstrand et al. \cite{Ekstrand2018} examined gender distribution of item recommendations in the book RS domain. Results prove that commonly deployed CF models differ in the gender distributions of generated item recommendation lists, such that neighbour-based approaches are shown to proportionality reflect user-item preferences in their reading histories, whereas model-based matrix factorisation favor books whose author is of male gender.
Furthermore, Ekstrand et al. in \cite{Ekstrand2018b} study the effect of recommendation algorithms on the utility for users of different gender groups, finding difference in effectiveness across gender groups.
Such work highlights that the effect in utility does not exclusively benefit large groups, implying that there may be other underlying latent factors that influence recommendation accuracy.
To address such issues of disproportionate gender treatment in recommendations, Edizel et al. in \cite{Edizel2019} have recently proposed a novel means of mitigating the derivation of sensitive features (such as gender) in the latent space, using fairness constraints based on the predictability of such features. A similar approach proposing fairness-aware tensor-based recommendation is also presented by Zhu et al. in \cite{Zhu2018}.

In the music domain, Aguiar et al. \cite{Aguiar2018} propose a methodology to assess the extent to which artists ranked in Spotify playlists are affected by gender after accounting for plausible determinants of inclusion on playlists such as country, song characteristics (e.g. bpm, key signature), and past streaming success.
The authors find that there is some evidence consistent with the presence of bias (both for and against female artists), however they do not draw subsequent relations between this and the disproportionate low streaming share of female artists on the platform.
In the work by Anglata-Tort et al. \cite{AngladaTort2019}, through the analysis of UK top 5 music charts between the years 1960-1995, authors show how popular music is affected by a large gender inequality, showing the presence of an existing bias in the listening preferences towards male artists.
Similarly, Millar in \cite{Millar2008}, surveying a population of Australian young adults, shows how music preferences are affected by gender bias, evidencing differences between male and female listeners. 
In contrast, in our work we apply an auditing strategy for bias propagation showing under which conditions input preferences are reflected in RS output, inferring music preferences from the users' listening history grouped with respect to the artists' gender.

\section{The LFM-1b Dataset}
\label{subsec:lfm1bdataset}
The LFM-1b dataset consists of more than one billion listening events created by over 120,000 users of the music streaming platform Last.fm \cite{Schedl2016}. 
In our analysis, we consider user-artist playcounts formed by aggregating user-song listening events by common artists. We then scale logarithmically the number of listens, as done in \cite{Jawaheer2010, Dean2020}. 
We work with a filtered version of the dataset in which:
a) we remove users who listened to less than 10 unique artists, and artists listened to by less than 10 users;
b) we discard users whose listening history contains more than 25\% of artists with unknown gender, to mitigate the impact of artists with missing gender in the dataset. 

User gender is represented in the dataset with three categories: \textit{male}, \textit{female} and \textit{N/A}.
We choose to focus only on users with self-declared gender, working with two final categories of user gender: male and female.
As shown in Table \ref{tab:recap}, distributions are highly imbalanced towards men --  72\% of the users are men. 

\begin{table}[h!]
 \begin{center}
  \begin{tabular}{ccccc}
    \toprule
    &\multicolumn{2}{c}{LFM-1b}&\multicolumn{2}{c}{LFM-360k} \\
    &\textit{male}&\textit{female}&\textit{male}&\textit{female} \\
    \midrule
    \makecell{Users\\ \textit{\%}} &\makecell{31.4K\\\textit{71.67}}&\makecell{11.5K\\\textit{28.33} } & \makecell{94.3K\\\textit{75.40}}&\makecell{30.8K\\\textit{24.60}} \\
    \hline
    \makecell{Artists\\ \textit{\%}} &\makecell{127K\\\textit{82.30}}&\makecell{27.3K\\\textit{17.70} } & \makecell{50.4K\\\textit{82.83}}&\makecell{10.5K\\\textit{17.17}} \\ 
    \hline
    \makecell{Top-head \\ \textit{\%}} &\makecell{25.7K\\\textit{85.21}}&\makecell{4.8K\\\textit{15.79} } & \makecell{10.1K\\\textit{86.99}}&\makecell{1.5K\\\textit{13.01}}  \\
    \hline
    \makecell{Long-tail \\ \textit{\%}} &\makecell{100K\\\textit{81.87}}&\makecell{22.2K\\\textit{18.13} } & \makecell{38.7K\\\textit{81.95}}&\makecell{8.5K\\\textit{18.05}} \\
    \bottomrule
  \end{tabular}
 \end{center}
  \caption{Users' and artists' distributions after the filtering process. ``Top-head'' artists are the top 20\% of artists by play counts, while the remaining 80\% are the ``long-tail.''}
  \label{tab:recap}
\end{table}


\begin{table}[ht!]
 \begin{center}
  \begin{tabular}{ccccc}
    \toprule 
     \multicolumn{5}{c}{LFM-1b} \\
     \midrule
    No. & Male artist& Plays& Female artist& Plays\\
    \midrule
    1&Radiohead&2.6M&Lana Del Rey&1.2M \\
    2&The Beatles&2.5M&Lady Gaga&1.1M \\
    3&Pink Floyd&2.1M&Rihanna&0.8M\\
    4&Daft Punk&2.0M&Bj\"{o}rk&0.7M\\
    5&Metallica&1.9M&Madonna&0.6M\\
     \midrule
    \multicolumn{5}{c}{LFM-360k} \\
    \midrule
    1&Radiohead&6.2M&Bj\"{o}rk&1.3M \\
    2&The Beatles&5.4M&Avril Lavigne&1.1M \\
    3&In Flames&4.9M&Madonna& 1.1M\\
    4&Metallica&4.3M&Britney Spears& 0.9M\\
    5&Muse&4.2M&Regina Spektor& 0.9M\\
    \bottomrule
  \end{tabular}
 \end{center}
  \caption{Top 5 artists ordered by total play counts in LFM-1b and LFM-360k datasets.}
  \label{tab:top5}
\end{table}

Artist gender is not represented in the LFM-1b dataset, consequently we retrieve this information from the open music encyclopedia MusicBrainz\footnote{\url{https://musicbrainz.org/}} (MB) \cite{Swartz2002}. Code repositories to implement the following approach are made openly available\footnote{{https://github.com/dshakes90/LFM-1b-MusicBrainz-Gender-Wrangler}} alongside the acquired results of the data wrangling\footnote{\url{https://zenodo.org/record/3964506\#.XyE5N0FKg5n}} to elicit reproducibility .

We identify five discrete categories of gender defined in the MB database: \textit{male}, \textit{female}, \textit{other}, \textit{N/A} and \textit{undef}. In the case of artists of gender \textit{N/A} and \textit{undef}, these are differentiated by artists for which gender is not applicable and identifiable respectively.
For bands, we compute gender counts of all members and then compute an overall classification based on whichever count has a majority. In the case of artists with gender ties (e.g, a band consisting of 2 males and 2 females), we discard such artists from our final analysis as gender is in this instance, deemed ambiguous.
After applying this methodology, we are able to identify 27\% of artists with a known-gender. Distributions are observed to be highly imbalanced such that artists of male gender consist of the majority (82\%) of artists for which gender can be identified, as shown in Table \ref{tab:recap}.

In our final analysis, we further filter artists not identified as male or female according to the procedure described above.
Artists of gender \textit{other} are discarded as we deem such data to be too sparse to be informative in the analysis of users' listening preferences. We note this group merits further future evaluation, perhaps relying on qualitative methods, and limitations of this binary approach are discussed in Section \ref{sec:conclusion}.
Table \ref{tab:top5} presents the top 5 artists based on the total sum of play counts in the filtered LFM-1b dataset. We observe a trend for male artists' popularity, having approximately twice as much play counts as top-rated female artists/bands. We also observe a trend for the top male artists on the platform to be more commonly composed of bands in comparison to the top-rated female artists.

\section{The LFM-360k Dataset}
\label{subsec:lfm360kdataset}

The LFM-360k dataset \cite{Celma2010} consists of approximately 360,000 users listening histories from Last.fm collected during Fall 2008, presenting a snapshot of listening activity for an earlier period in comparison to the LFM-1b dataset.
With respect to user gender distributions the proportion of users with a self-declared gender rises to 91\% whereas similarly to the LFM-1b dataset, artist gender is not defined.
To resolve this, we implement the same pre-processing methodology with the MB database as described for the LFM-1b dataset.
After further applying the filtering criteria previously detailed, we are able to identify 31\% of artists with a known gender, a proportion notably higher than that of what we were able to identify for the LFM-1b dataset.
As presented in Table \ref{tab:recap}, artist gender distributions in the filtered dataset are once again highly imbalanced towards artists classified as men. For users with identified gender, we again observe a high imbalance towards male users (75\%) comparable to rates observed in the LFM-1b dataset. 
When comparing the two datasets we observe several additional differences and similarities which may impact the propagation of a gender bias in artist recommendations.
First, the number of users is significantly larger than that of the LFM-1b, whilst the number of artists is much smaller.
Second, sparsity is higher in the LFM-360k dataset in comparison to the LFM-1b.
Third, with regard to the top 5 artists of male and female gender in the dataset we observe significantly higher playcounts for artists classified as male in comparison to the LFM-1b dataset, as shown in Table \ref{tab:recap}.
With regard to similarities across the two datasets, we observe that top 5 popular male artists are more commonly bands in comparison to the top 5 female artists.
In addition, we observe that the long-tail of both datasets contains significantly higher distribution of female artists, in comparison to the top head reinforcing the conclusion that female artists are significantly more likely to be less popular on the Last.fm platform and hence, more likely to be less recommended as a result of this popularity bias. 

\section{Methodology}
\label{sec:methodology}
\subsection{Evaluation Metrics}
\label{subsec:bias}
In this section, we formally outline the metrics of preference ratio, bias disparity, as well as accuracy and beyond-accuracy metrics considered during the evaluation.

\textbf{Preference ratio (PR)}. Let $U$ be the set of $n$ users, $I$ be the set of $m$ items and $S$ be the $n$x$m$ input matrix, where $S(u,i) = 1$ if user $u$ has selected item $i$, and zero otherwise. Given matrix $S$, the input preference ratio for user group $G$ on item category $C$ is the fraction of liked items by group $G$ in category $C$, formally defined as the following:
\begin{equation}
    P R_{S}(G, C)=\frac{\sum_{u \in G} \sum_{i \in C} S(u, i)}{\sum_{u \in G} \sum_{i \in I} S(u, i)}
\end{equation} 

\textbf{Bias disparity (BD)}. It is defined to be the relative difference between the preference bias for input $S$ and output of a recommendation algorithm $R$. Formally we define the metric as the following: 
\begin{equation}
    B D(G, C)=\frac{P R_{R}(G, C)-P R_{S}(G, C)}{P R_{S}(G, C)}
\end{equation}

In our analysis, we generate a set of $r$ ranked items, $R_u$ which have the highest predicted ratings for a given user $u$, limiting the value of $r$ to 5. 

\textbf{Accuracy and beyond-accuracy metrics}. To evaluate the RS performance, we additionally deploy two accuracy metrics: \textit{Precision}, \textit{nDCG}, and three beyond-accuracy metrics: \textit{coverage}, \textit{spread} and \textit{long-tail percentage}. We refer to the metrics formulation as detailed in the work by Noia et al. \cite{DiNoia2017}. 
\textit{Precision} (\textit{p@n}) captures the proportion of relevant items in top-N recommendations, such that relevance is a binary function that represents the relevance of item $i$ for a user $u$. In our work, we consider relevant a recommendation which is greater or equal to the average scaled listening count for a user, after discarding outliers in the data computed using the interquartile range. Although \textit{p@n} is useful for analysing generated item recommendations, it does not capture accuracy aspects relating to the rank of a recommendation. Hence, in our work we also deploy the metric \textit{nDCG}, a rank sensitive metric used to evaluate the accuracy of a RS. 
With respect to metrics beyond accuracy, we utilise both \textit{spread} and \textit{coverage} to capture a recommender systems ability to recommend a broad range of unique items. Such approaches are important to consider in our work to potentially reason and explain bias propagation across artist genders. 
The metric \textit{long-tail percentage} is used to capture the proportion of item recommendations which exist in the long tail. In our work, we define the long tail as the 80\% of least popular items in the system. We use the metric to capture a filtering algorithms capacity to display the popularity bias.

\subsection{Recommendation Algorithms}
\label{subsec:algorithm}

We test several commonly deployed memory- and model-based CF algorithms, following a similar approach to previous work \cite{Kowald2020,Lin2019}. 
Using Surprise \cite{Surprise}, a Python library for recommender systems, we formulate our music recommendations as a rating prediction problem where we predict the preference of a target user \textit{u} for a target artist \textit{a}.
We then evaluate RS recommending the top-5 artists with the highest predicted preferences.

We consider two types of CF algorithms:
(1) KNN-based approach: \textit{UserKNNAvg} \cite{Yehuda2010}, and
(2) factorisation-based approach: \textit{Non-Negative Matrix Factorization} (\textit{NMF}) \cite{Luo2014}.
Hyperparameters of \textit{UserKNNAvg} and \textit{NMF} are tuned to give the best performance we can achieve with respect to the rank aware metric, \textit{nDCG}. 
In addition, we consider two \textit{MostPopular} and \textit{UserItemAvg} algorithms which respectively, recommend the most popular and highest rated artists. We consider these algorithms for a baseline comparison. 
%

A variation of the \textit{leave-l-out} evaluation detailed in \cite{Canamares} is performed whereby we translate the approach to evaluate a \textit{top-n} RS. 
Drawing influence from the methodology of Said et al. \cite{Said2013ATR} we define 3 parameters: (1) $n$, the size of the recommendation list generated, (2) $N$, the number of items selected for each user to appear in the test set. $N$ is constrained to be > $n$  to allow for variance in item recommendations across tested algorithms. (3) $M$, the minimum number of unique artists listened to by a user. $M$ is constrained to be > $N$ to ensure a non-empty test set is able to be formed for each user. 
We construct three folds, randomly selecting for each user, $N$ items in their listening history to belong to the fold's test set and then subsequently removing these listening events from the folds training set. 
For each of the algorithms tested, we compute all evaluation metrics and preference ratios over each fold and then subsequently report average performance. In our work we set $N$ = 10, $M$ = 20 and $n$ = 5, thereby generating top-5 recommendation lists. We consider a user's test set of size $N$ as the sample space for recommendations to be formed. 
%

\subsection{Experimental Design}
\label{subsec:design}

We set up two experimental designs to evaluate variations in gender bias disparity across recommended artists and user groups for the two datasets. For all experiments detailed, code repositories are made openly available\footnote{https://github.com/dshakes90/Last-fm-Gender-Bias-Analysis}.
Experiment 1 is a real-world scenario in which male and female gender distributions are representative of those in both datasets.
Experiment 2 is an extreme scenario in which all users have high levels of preference ratio, representing extreme listening preferences towards artists of a specific gender.

\textbf{Experiment 1}. We generate recommendations for a sample of all users for which gender can be identified. In the LFM-1b dataset, we limit the size of this sample to be 30\% randomly chosen of all male and female users in the whole dataset (approx 12,000 users), due to computational constraints.
The size of the user sample for the LFM-360k dataset was also constrained to be approximately the same size as samples for the LFM-1b dataset.
User and artist gender distributions in both samples are representative of overall gender distributions in the entirety of both datasets.
We therefore use this experiment to consider the case of gender bias propagation under a real world scenario, assessing the extent to which gender bias disparity may differ across datasets.

\textbf{Experiment 2}. We generate recommendations only for a sample of male and female users which have high preference ratios in the dataset, thereby simulating an extreme scenario under which all users are highly biased towards one artist gender group in their listening preferences.
For the LFM-1b dataset, we select the top 30\% of both male and female user groups with the highest maximum input preference ratios, maintaining both the proportions of male and female users in the datasets, and the sample size of experiment 1.
For the LFM-360k dataset, we sample users from both male and female user groups maintaining the distribution of male and female users in the original dataset. The final user sample has approximately the same sample size as that of the LFM-1b user sample.

\begin{figure}
\centering
   \includegraphics[width=1\linewidth]{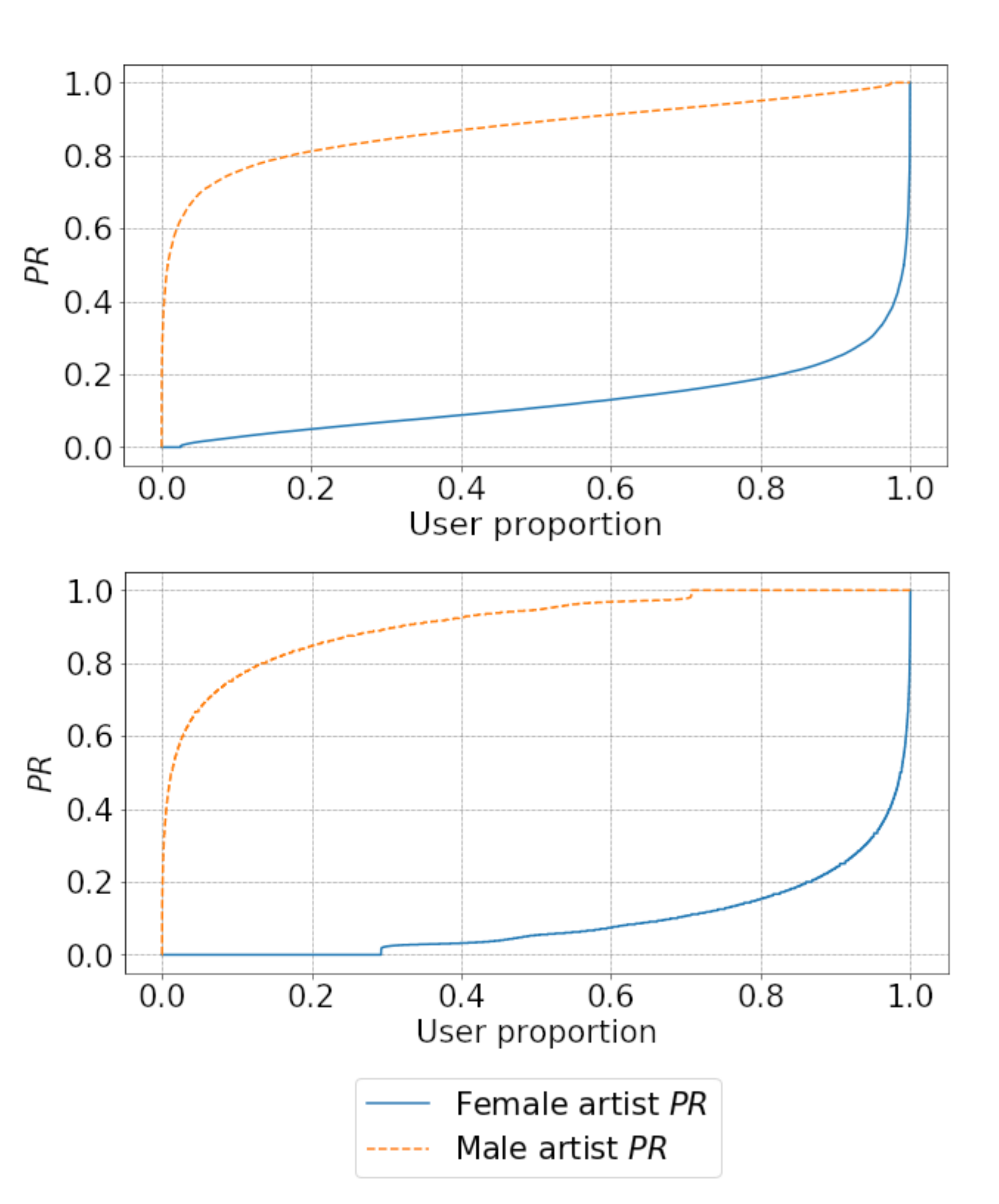}
   \caption{Input Preference Ratio (PR) distributions: LFM-1b (\textit{top}) and LFM-360k (\textit{bottom}).}
   \label{fig:distr} 
\end{figure}

Figure \ref{fig:distr} represents the distributions of users' input preference ratio towards male and female artist groups.
For both datasets considered in this study, it shows that only around 20\% of users have a preference ratio towards male artists lower than 0.8. On the contrary, 80\% of users have a preference ratio lower than 0.2 towards female artists.
Due to the disproportionate amount of users with extreme preferences for male artists across both datasets, a random sampling methodology proposed does little to assess extreme preference towards female artists, resulting in a situation very similar to experiment 1.
To resolve this, we further limit our sample space to only users who have extreme preference for female artists, with input preference ratio towards female artists $> 0.6$. This results in a sample size reduction to 100 users for the LFM-1b dataset, and 400 users for the LFM-360k dataset.
Although reduced in size in comparison to experiments 1, we believe such experimental designs to be fundamental to measure the extent to which the treatment of users with extreme preferences differs across artist genders.
Experiment 2 represents a situation opposite to the one proposed in experiment 1, thanks to which we can assess if bias propagation is not embedded in the gender \textit{per se}, but is a result of pre-existing bias.

\section{Results}
\label{sec:results}

\subsection{Experiment 1 - Whole population}
\begin{figure*}[ht!]
\centering
   \includegraphics[width=1.01\textwidth]{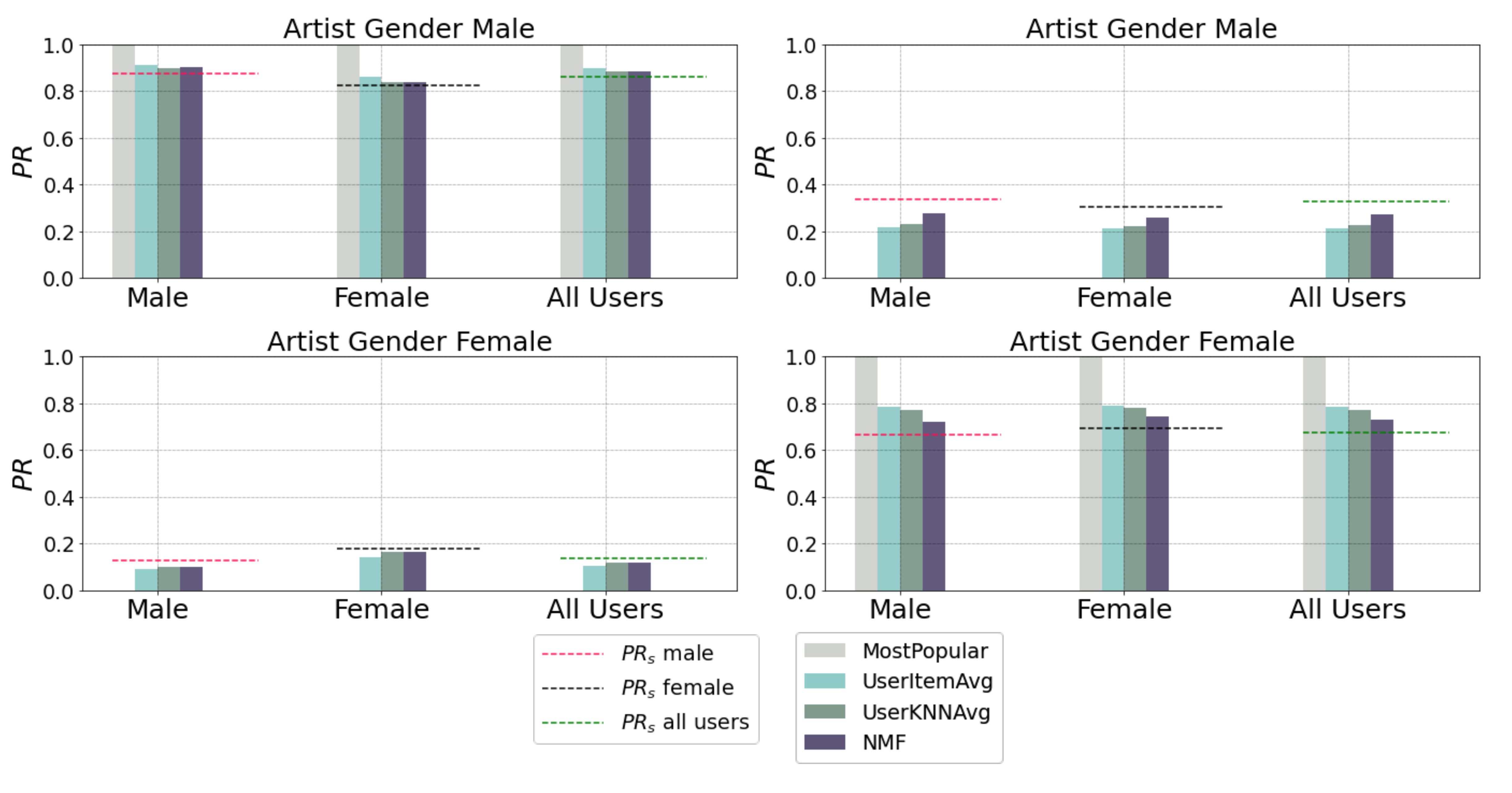}
   \caption{Preference Ratio (PR) results for LFM-1b dataset for experiment 1 (\textit{left column}), and experiment 2 (\textit{right column}).}
   \label{fig:pr_db} 
\end{figure*}

\begin{figure*}[ht!]
\centering
   \includegraphics[width=1.03\textwidth]{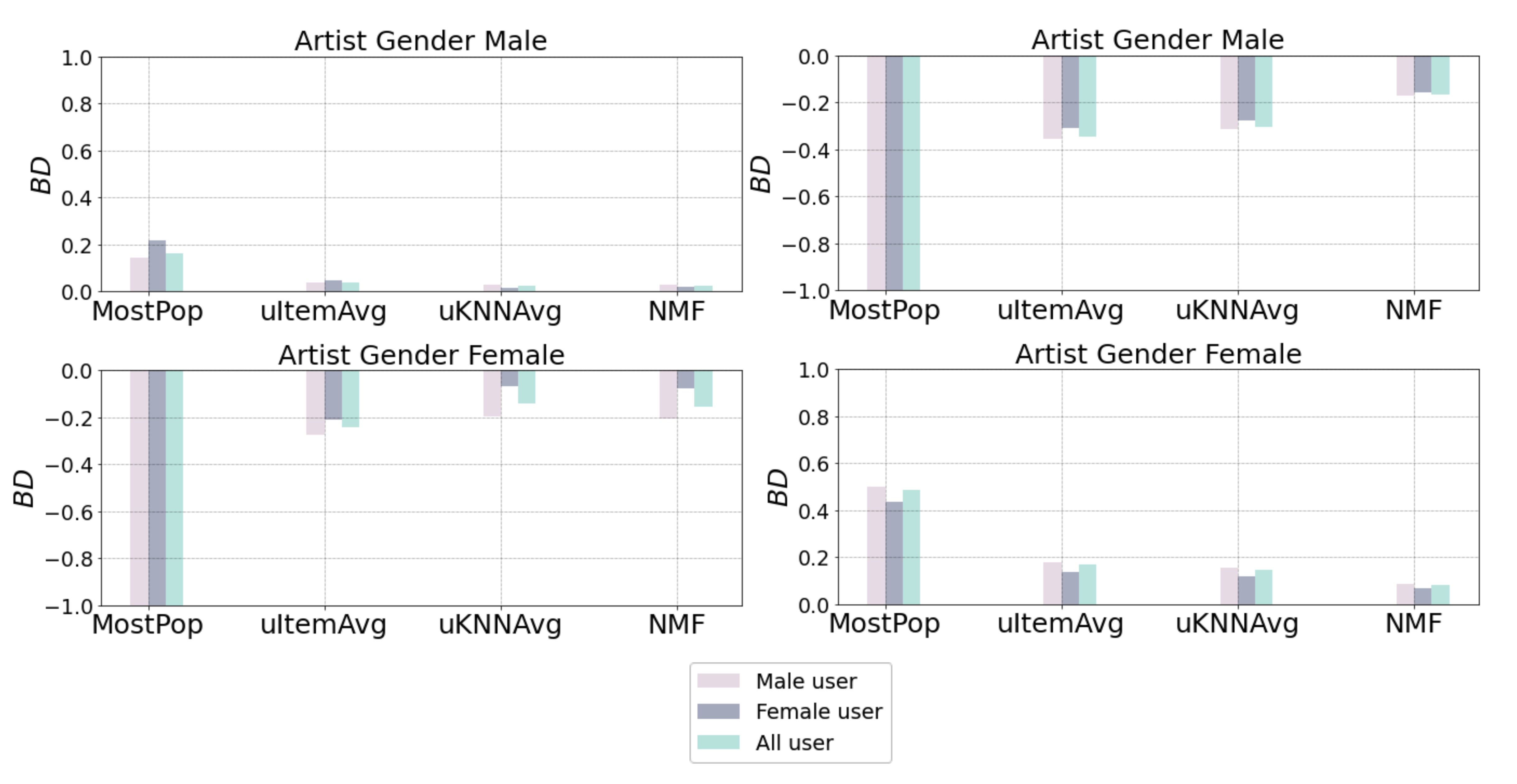}
   \caption{Bias Disparity (BD) results for LFM-1b dataset for experiment 1 (\textit{left column}), and experiment 2 (\textit{right column}).}
   \label{fig:bd_db} 
\end{figure*}

\begin{figure*}[ht!]
\centering
   \includegraphics[width=1.03\textwidth]{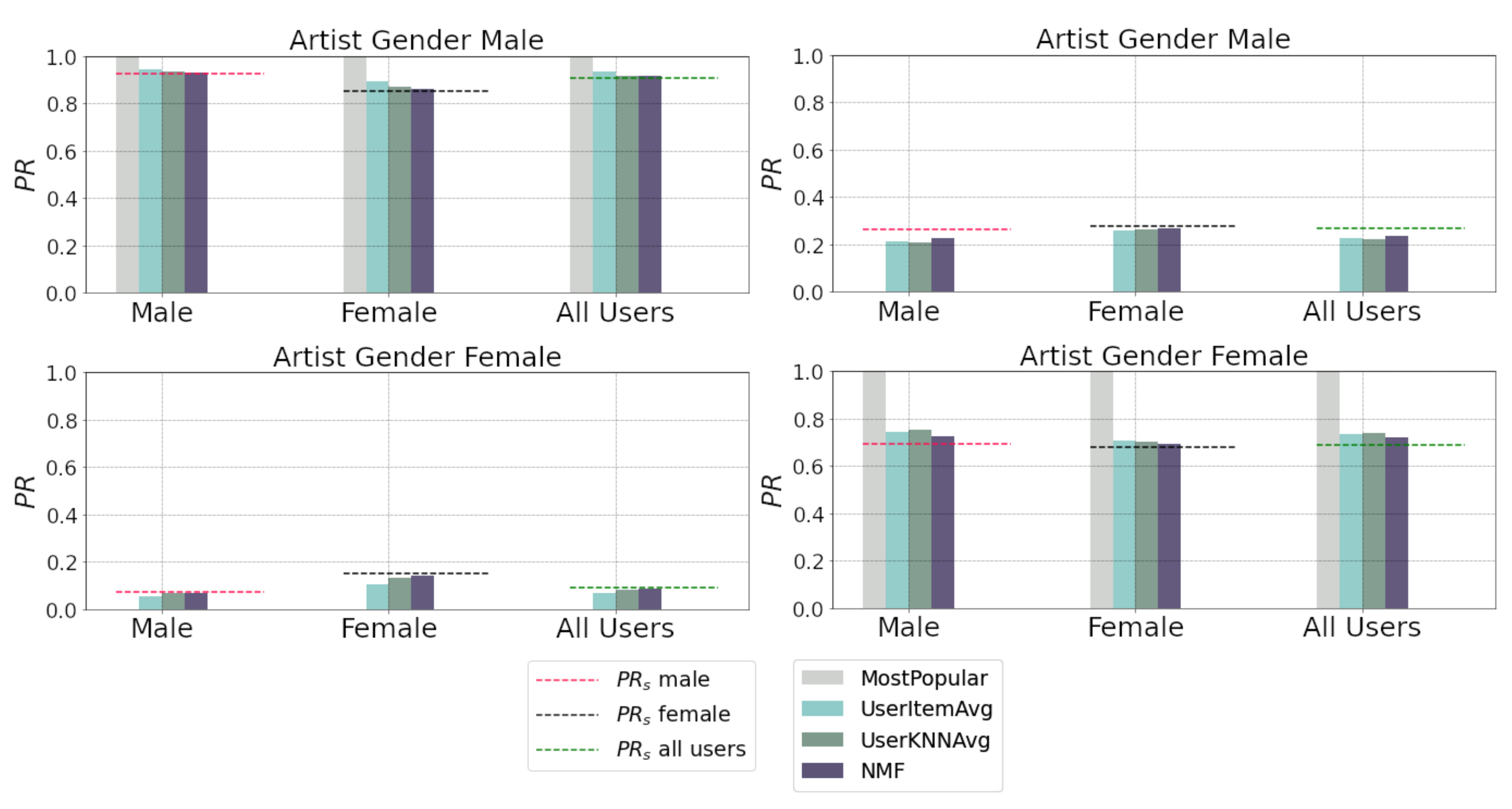}
   \caption{Preference Ratio (PR) results for LFM-360k dataset for experiment 1 (\textit{left column}), and experiment 2 (\textit{right column}).}
   \label{fig:pr2_db} 
\end{figure*}

\begin{figure*}[ht!]
\centering
   \includegraphics[width=1.01\textwidth]{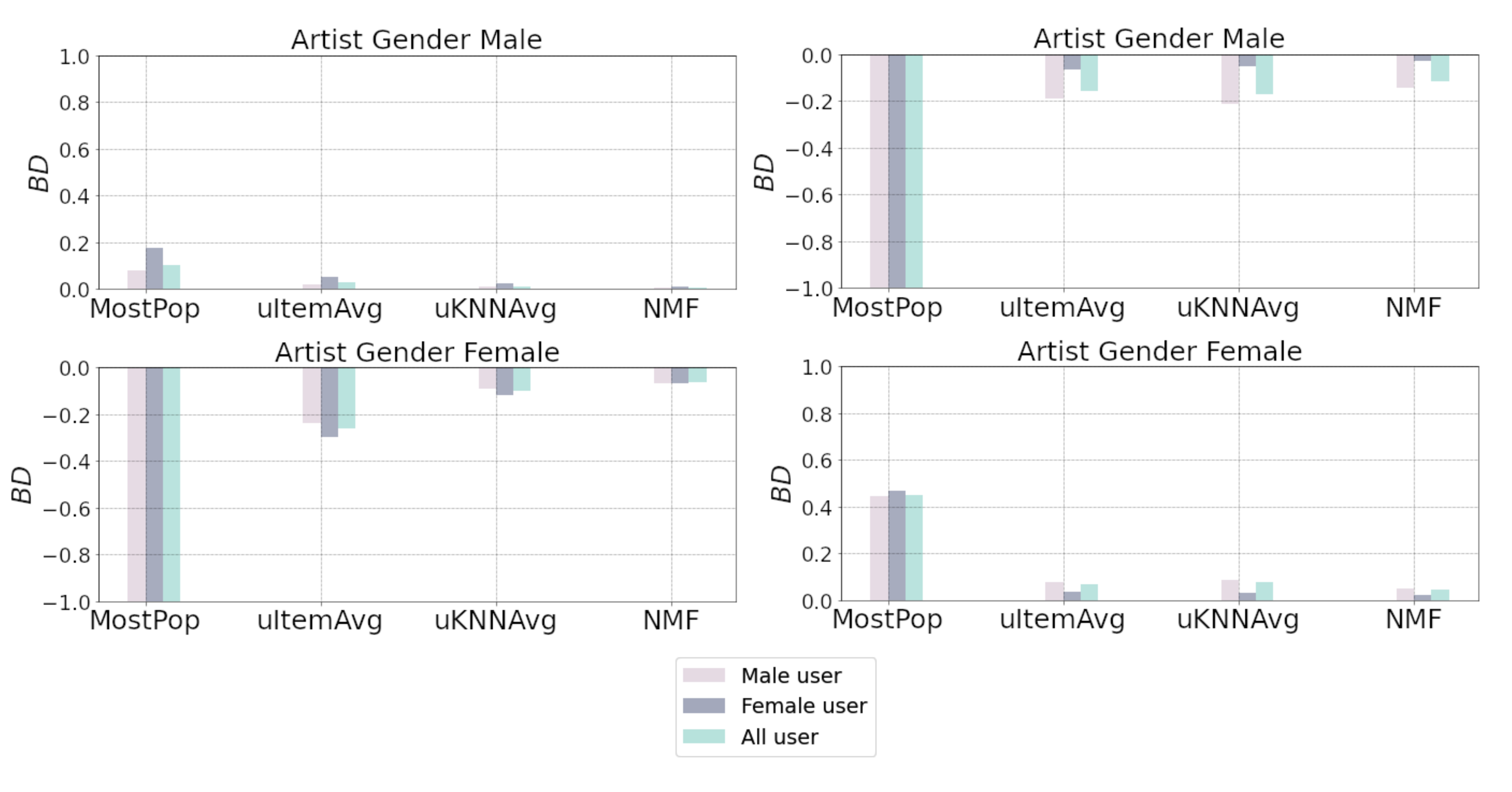}
   \caption{Bias Disparity (BD) results for LFM-360k dataset for experiment 1 (\textit{left column}), and experiment 2 (\textit{right column}).}
   \label{fig:bd2_db} 
\end{figure*}

We report in Figure \ref{fig:pr_db} preference ratio, and in Figure \ref{fig:bd_db} bias disparity results obtained with the LFM-1b dataset. Figure \ref{fig:pr2_db} and Figure \ref{fig:bd2_db} present preference ratio and bias disparity results respectively for the LFM-360K dataset.
The dotted lines in Figure \ref{fig:pr_db} and Figure \ref{fig:pr2_db} represent input preference ratios whereas the plot's bars display output preference ratios computed from generated recommendation lists. 
With regard to pre-existing bias, users in both datasets display high and low input preference ratios for male and female artists respectively, thereby in line with the findings of Millar~\cite{Millar2008}.
In addition, for both artist genders input preference ratios can be seen to be higher by users who share the same gender as the artist. 
With regard to bias propagation after recommendation, all recommendation models tested result in a positive bias disparity for male artists for which there is minimal variance in treatment across user genders. 
The popularity-based algorithm results in the highest levels of bias disparity for both male and female users, whilst the \textit{NMF} and \textit{UserKNNAvg} algorithms tested result in the lowest absolute levels of bias disparity with marginal difference in bias propagation across the two algorithms. Whatsmore, our findings show male users to be more affected by bias propagation in the LFM-1b dataset whilst for LFM-360K, we observe bias propagation to be greater for female users thereby inline with the findings of Lin et al. \cite{Lin2019}.
With regard to bias disparity for female artists, negative levels are observed for all algorithms tested. The \textit{MostPopular} algorithm results in the lowest levels of bias disparity due to female artists having significantly lower popularity for both datasets tested, as shown in Table \ref{tab:recap}. 
%
We observe bias propagation to be greater for recommendations generated using the LFM-1b dataset reflected in the lower \textit{long-tail percentage} attained. This suggests that users in the LFM-1b dataset may be more subject to a popularity bias in comparison to LFM-360k which may translate to increased levels of gender bias disparity due to female artists proportionally residing less in the top-head. 
Together, our findings suggest that differences in bias propagation across the two datasets may be traced to pre-existing bias entering the system in the form of listening events. 


\begin{table}
 \begin{center}
  \begin{tabular}{lcccc}
    \toprule 
    & \makecell{\textit{Most} \textit{Popular}} & \makecell{\textit{UserItem} \textit{Avg} }& \makecell{\textit{UserKNN} \textit{Avg}}& \textit{NMF} \\
    \midrule
    precision & 0.010 & 0.595 & 0.676 & \textbf{*0.734} \\
    nDCG & 0.012 & 0.663 & 0.793 & \textbf{*0.880} \\
    coverage & 1.7E-04 & 0.364 & \textbf{*0.558} & 0.552 \\
    spread & 2.322 & 11.85 &  \textbf{*12.84} & 12.72 \\
    longtail \% & 0 & 0.027 & 0.053 & \textbf{*0.054} \\
    \bottomrule
  \end{tabular}
 \end{center}
  \caption{Experiment 1 evaluation results on the LFM-1b dataset. Values in bold represent the top value, while marked with * are results where the difference is statistically significant, according to a t-test with $\alpha = 0.05$.}
  \label{tab:results}
\end{table}

\subsection{Experiment 2 - Extreme preferences}

Considering users with extreme preferences for female artists we observe the inverse scenario of experiment 1, such that bias disparity is positive for female artists and negative towards male artists, as shown in Figure \ref{fig:bd_db} and Figure \ref{fig:bd2_db}.
For both datasets, we comment that one cause of such disparity is a dramatic imbalance in users' listening preference, which then subsequently propagates through to other users' recommendations.
Our findings show that such bias propagation is not reserved for male artists on the platform and can, under extreme scenarios emerge in the opposite manner.
For both memory- and model-based approaches tested we observe significant differences in bias disparity: \textit{NMF} results in the smallest absolute bias disparity increase thereby reflecting a users' input preference, whereas the neighbour-based \textit{UserKNNAvg} increases absolute bias disparity levels towards whichever user-artist preference is in the majority. 
The tendency of \textit{NMF} to propagate less bias, positively or negatively speaking, in comparison to the other models is also reflected in the results obtained from the beyond-accuracy metrics evaluation.
Indeed, for experiment 2 \textit{NMF} achieves the high levels of coverage, recommending wider subsets of artists, and at the same time high levels of recommendation spread.
Together these results suggest that the model-based algorithm considered in this study is capable of achieving a higher level of diversification in the outcomes in comparison to the memory-based model.
Translated to our scenario, it means that \textit{NMF} is the algorithm that focuses less on recommending a specific gender group, avoiding the exacerbation of pre-existing bias in the dataset that other recommendation algorithms exhibit. 
Again, the effect of bias propagation is seen to be more amplified in the case of the LFM-1b dataset.

\section{Conclusions and Future Work}
\label{sec:conclusion}

Studies of gender bias in music preferences, conducted in a field such as Music Psychology and Gender Studies, have already evidenced how socio-cultural factors are responsible for disparate treatment of not-male artists.
In the field of MIR, relatively little research has analyzed how existing technology can have a role in mitigating or amplifying this bias.
In line with the studies on bias disparity in the RS literature, focusing on the musical domain we show how recommendation outcomes can actually impact gender bias in music preferences. 
Using a binary gender classification, where users and artists are classified as male or female, we have shown how at different levels recommender systems can propagate a pre-existing bias.
In addition, simulating an ``upside down'' world where users have a much higher preference towards female artists, still we find evidence of an exacerbation of that bias. Our results show that gender bias can be propagated by CF-based recommendations, according to the bias present in the data.
Hence, RS can have a role in propagating bias, but at least in our exploratory study, we have not found evidence about if they cause the emergence of new forms of biases.

The limitations of our work are several.
First, it is important to remark that the binary classification of gender is an oversimplification of gender representation. 
The state of the art perspective of gender from both natural and social science domains is often non-binary, where male and female are just one of the many genders in which an individual may choose to identify by. 
Binary definitions of gender have been widely critiqued to be socially constructed through routine gendered performances \cite{Beauvoir2015, butler_2006} thereby, considering gender to be only binary in this work is both limiting and to some degree, reinforcing of such binary logic.
%
Second, the evaluation of RS is computed such that the impact of the outcome can be intended in the short- but not in the long-term. 
Using longitudinal data or simulation frameworks, we believe that a better comprehension of the phenomenon can be achieved, complementing the results we have presented.
Lastly, Last.fm users tend to come mostly from Western countries, consequently our results cannot be generalized to represent a global scenario. 
This issue is well known in the MIR domain~\cite{Serra2013}, and we do believe that to consider a multicultural perspective is undoubtedly a necessary step to give robustness to MIR studies dealing with socio-cultural and socio-technical phenomena.

\section{Acknowledgments} This work is partially supported by the European Commission under the TROMPA project (H2020 770376).

\bibliographystyle{ACM-Reference-Format}
\bibliography{main}

\end{document}